\documentclass[graybox]{svmult}

\usepackage{mathptmx}       
\usepackage{helvet}         
\usepackage{courier}        
\usepackage{type1cm}        
\usepackage{makeidx}         
\usepackage{graphicx}        
\usepackage{multicol}        
\usepackage[bottom]{footmisc}

\makeindex             

\begin{document}

\title*{Spreading in Social Systems: Reflections}
\author{Lehmann and Ahn}
\institute{Sune Lehmann \at Technical University of Denmark, DK-2800 Kgs.~Lyngby, \email{sljo@dtu.dk}
\and Yong-Yeol Ahn \at Indiana University, Bloomington, IN 47408, USA, \email{yyahn@iu.edu}}
%
%
\maketitle

\abstract{In this final chapter, we consider the state-of-the-art for spreading in social systems and discuss the future of the field. 
As part of this reflection, we identify a set of key challenges ahead. 
The challenges include the following questions: how can we improve the quality, quantity, extent, and accessibility of datasets? How can we  extract more information from limited datasets? How can we take individual cognition and decision making processes into account? How can we incorporate other complexity of the real contagion processes? Finally, how can we translate research into positive real-world impact? 
In the following, we provide more context for each of these open questions.}

\section*{Introduction}

As a starting point, we believe that social contagion will play a key role in shaping how society and democracy develops in the coming decades. 
As our world has become increasingly connected through the networks of social media, the role of social contagion has grown. 
Social media services, such as Twitter, Facebook, or Reddit are becoming the main channels through which people communicate and consume news. 
Because of these platforms' global connectedness, a piece of news---fake or not---can spread to millions of people around the world at near instantaneous speed. 
Moreover, the increasing social media use, combined with sophisticated machine learning algorithms for content recommendation, means that we increasingly find ourselves within comfortable ideological bubbles. 
Inside each bubble, content that reinforces our beliefs and biases will spread more easily among people with shared ideologies and potentially entrench people. 
Such entrenchment may grow in the future. 
Thus, humanity's major challenges are beginning to revolve less around building the right technologies, but more around puncturing bubbles in order to reduce societal polarization. 

The power to manipulate and control people's beliefs through social contagion is a double-edged sword. 
Such power can be used for public good---to effectively spread informed opinions on public health matters: safe sex, smoking, or vaccination to name a few examples. 
At the same time, however, this power can be, and has been, misused for manipulating public opinions or influencing the outcome of elections.
We expect that the impact of social contagion on our society---particularly on the foundation of democracy---will keep increasing.  
The social responsibility of research into social contagion processes should not be overlooked. 
\begin{center}
	$\bullet\,\bullet\,\bullet$
\end{center}
As is clear from the fantastic contributions in this book, our understanding of social spreading processes has advanced significantly over the past decade.  
Still, there are of course outstanding challenges. Among many, here we discuss the following:
\begin{itemize}
\item How can we improve the quality, quantity, extent, and accessibility of datasets? 
\item How can we extract more information from limited datasets? 
\item How can we take individual cognition and decision making processes into account? 
\item How can we incorporate other complexities from the real contagion processes? 
\item How can we translate research into positive real-world impact?
\end{itemize}

\section*{Please sir, I want some more data}
History tells us that the availability of high-quality data is a key driving force in science. 
The science of social contagion is no exception. 
In the past decade, datasets from long-term longitudinal studies, such as the Framingham Heart Study, as well as other massive online social media datasets have been the main fuel source that propels the study of social contagion. 
So, a natural question is `how can we get our hands on better datasets?' 

We should probably begin by asking what `better' data would even mean. 
`Better' may mean simply more details and larger volume. 
For instance, high-resolution data can reveal insights that are completely hidden when that same dataset is aggregated. 
Larger datasets imply increased statistical power and the ability to identify minute effects. 
Having more attributes can lead to the discovery of new associations or more precise control of confounding factors. 

Going beyond size and detail, better data may also imply a shift from \textit{found} data to more \textit{designed} data~\cite{salganik2017bit}.
Instead of re-purposing observational datasets, one can specifically design a (controlled) experiment and collect data. 
To do so, one should either create one's own data collection environment (e.g.~Sensible DTU \cite{stopczynski2014measuring}) or leverage existing services (e.g.~controlled experiments conducted by Facebook \cite{kramer2014experimental, bond201261}). 
The former is more constrained by resources and difficult to scale, while the latter is more constrained by the economic incentives of the company and details of the services. 
Collaboration between academia and industry is a nice hybrid approach and has produced many successful insights (see for example \textbf{Part 4, Chapter 3}). 

Finally, we stress the importance of open access to data. 
Even if one collects an ideal dataset to study social contagion, the dataset may make little impact on the field if the data cannot be shared with others. 
Even the validity of a study that uses this ideal data may not be ensured if no one can use the data to replicate the results. 
The benefits of data sharing are clear; making a dataset public can maximize the impact of the dataset and makes the resulting research more transparent and reproducible. 

However, many social datasets are not easy to share in a raw form (or even collect) due to privacy concerns. 
There have been several efforts from industry to share anonymized datasets but many have unfortunately failed. 
For instance, Netflix shared a large dataset for a highly profiled recommendation engine challenge, only to find that the dataset could be easily de-anonymized~\cite{narayanan2008robust}. 
Later, an anonymized Flickr social network was de-anonymized using Twitter's social network~\cite{narayanan2011link}. 
After several incidents of this kind and more theoretical developments, it has become clear that it is very difficult to properly anonymize data, in particular data that involves social networks. 

There can also be other kinds of backlash related to sharing data, or simply just sharing results of studies that are conducted in industry. 
For instance, the emotional contagion study published by Facebook in collaboration with academic researchers \cite{kramer2014experimental} upset many users and put Facebook in a difficult position. 
The adverse reaction to this study may have suppressed the in-house research efforts across industry and reduced incentives to publish academic articles, not to mention datasets. 
Thus, understandably, most companies are cautious about sharing raw datasets and even results of their internal experiments. 

At the same time, the push from publishers, scientists, and other advocates for open data has begun to produce practical solutions. 
These practical solutions aim for a compromise between level of detail within the data and privacy converns. 
A common approach is to share data that is sufficiently aggregated so that the re-identification or extraction of any individual data is impossible. 
Another solution is to maintain a special internal repository for replication data as well as mechanisms for external researchers to access the data upon request. 
Such solutions may address the issue of replicability, but fall short with respect to replicating the full benefits of open datasets. 
It will be interesting to see whether it will become easier to access raw datasets from industry through improved privacy-conserving algorithms or whether we will see aligned efforts resulting academia-industry collaboration in the future. 

\section*{Homophily or contagion?}
Although we now have unprecedented amounts of data related to social contagion---and describing social behavior in general, most available datasets are still observational. 
This fact imposes serious limitations. 
A central issue is that, because of homophily (and latent homophily) in networks, it is difficult to perform causal inference. 
As the heated debate regarding the series of papers using the Framingham Heart Study---an observational dataset---has demonstrated \cite{christakis2007spread, fowler2008dynamic, christakis2008collective, cacioppo2009alone, aral2009distinguishing, shalizi2011homophily, lyons2011spread, christakis2013social}, causal inference based on observational data is a major challenge, and the effort to extract as much as information from observational datasets will continue.  
A number of methods have been developed to more clearly understand the limitations and extract more information from observational datasets (see \textbf{Chapter 3}). 
The results from observational data will remain as an important part of the social contagion research. 

\section*{Micro-contagionomics}
Most existing studies assume fairly simple contagion models that do not take into account complex individual decision making and variations across individuals. 
Given what is known about cognition and social psychology, another interesting avenue of research will be to incorporate cognitive and psychological models of decision-making and behavioral changes into the study of social contagion; both in theoretical and empirical studies. 

Although there are many theoretical models, rich models that can capture more nuanced cognitive limitations and biases---such as complex interactions between beliefs~\cite{rodriguez2016collective} or limited attention~\cite{weng2012competition}---as well as the nature of contagion~\cite{weng2013virality} will be needed to fully understand and better model social contagions. 
%
%

On the empirical front, we need more precisely controlled, high-resolution experiments. 
In spite of all the progress there has been made studying empirical patterns of information diffusion (\textbf{Part 3}), we are still limited to examining overall patterns and the results of spreading. 
In fact, outside of purely theoretical models (\textbf{Part 2}), we have little idea how to incorporate knowledge and insights from psychology and cognitive science in order to measure the microscopic mechanisms that govern the adoption of a new idea.

Within the empirical work, we mostly study \emph{proxies} for the information that is truly spreading, whereas the work on random control trials (\textbf{Part~4}) focuses on observing \emph{behaviors} resulting from a spreading process on an underlying network.
Thus, a possible way forward could be through new experimental paradigms, where we study both the spreading agent on its journey through the network, along with well-defined behavioral changes on the individual-level. 

To make this concrete, let us outline some thought experiments. 
An extreme one will be similar to a reality show, where every single conversation and related behavior is recorded \cite{roy2002learning}, with added potential interventions and controls. 
The data then could then be analyzed to identify how exactly the information spread through the participants. 

Another possible experimental design would begin with designing specific, well defined pieces of information designed to illicit a reaction (or lack thereof) that can be measured (e.g.~going to collect free beer at a certain location, pressing a certain button). 
Further, study participants must only be able to access these pieces of information in a way that reveals the identity of the person in question (e.g.~by displaying this information on a personalized web-page or via a mobile-phone app). 
Finally, of course, information on how to access these pieces of information must travel in a well-defined way on the social graph independently of the communication platform (email, online social network, face-to-face). 
While accessing the piece of information, we could also provide information about actions (or information state) of the network neighbors.
Starting from randomized control trials and with access to both detailed spreading paths on the network as well as behavioral outcomes, such an experimental paradigm would allow us to begin collect reliable statistics and answer questions on the microscopic mechanisms that shape spreading and adoption, such as how the probability of spreading depends on the local network structure.

By running multiple experiments we would also be able to empirically examine the role of `stickyness' or `sexyness' of ideas in spreading, acknowledging that intrinsic properties of the spreading agent might interact with the network in a non-trivial way. 

\section*{It's complicated: multi-layered, dynamic, co-evolving networks}
Over the past twenty years or so, we have made substantial progress in our ability to describe and analyze static complex networks. 
But real networks exhibit many complex features. 
For instance, networks change dramatically over time.  
The connection patterns of social networks are constantly reconfigured as we connect with friends, co-workers, and family---as we move through our daily lives, as we adopt new platforms for communication. 
Our theoretical foundation for analyzing and understanding temporal networks is solidifying, but we are still learning how to treat the interplay between temporal networks and the dynamics of network spreading on those networks. 

Network structure is not just changing in isolation. 
Often the dynamic evolution of a network is due to the social contagion in the network. 
In other words, the structure of social network and the dynamics of social contagion co-evolve \cite{snijders2007modeling, gross2008adaptive}. 
We can re-examine the issue of homophily versus contagion in this context.  
It is not just that these two concepts are confounded (a difficult problem in its own right).  
It is also that reality is often a mixture of the two (an even more difficult problem in its own right). 
In the wild, we are likely to see a dynamic bidirectional interplay of influence and homophily on each dyad---and more generally within each network neighborhood---shaping the evolution of the network itself, as well as the dynamics of information flowing through it.

%
%
While the advent of online social media and other communication channels have opened up new ways to study society and our communication patters, online social media have also had the less publicized effect of increasingly fragmenting social communication across multiple channels. 
Most people use multiple social media services, often each for different purposes. 
As the main communication channel for their friends, some may use Facebook, some may use Twitter, some may use Snapchat, and some others may not even use any social media services at all, using only `traditional' channels such as in-person conversations. 
Thus, even when some company dominates in many markets across the globe, a single service only captures a small, biased fraction of threads in this fabric of social communication. 
For instance, a single instance of social contagion may manifest itself as numerous disjoint spreading events if observed through the lens of a single service. 
And thanks to homophily and network effects, users of a service, and the ways that they use the service, tend not to be random samples from the full population. 

An important implication in terms of the study of social contagion, is that even the largest studies, if they were conducted on a single platform, might be lacking significant spreading events that occur via other channels and making conclusions based on biased behavioral patterns. 
Thus, it is important to ask: how can we know that the observed results are not artifacts of such a fragmentation?  
How can we study spreading phenomena that occur across many communication channels? 

We believe that progress on this topic will occur by working simultaneously on both the theoretical and empirical side, with each side complementing the other.  
Empirical observations describe how people juggle multiple types of social media and how information spreads across the different layers of social networks will provide good insights on how to model use of the fragmented networks. 
Theoretical studies on multi-layer information spreading processes will then inform hypotheses and suggest general patterns to be tested through additional empirical studies. 
Collaborations across multiple social media platforms are also needed to obtain proper datasets to study multi-layer diffusion. 
Smaller-scale, but higher-resolution studies also have great potential to deepen our understanding of how people use multi-layered social fabric. 
Finally, because it will be practically impossible to capture every possible social interaction, statistical inference techniques and theoretical studies to understand the effect of missing data---or missing layers---and to infer the missing data will be necessary.

\section*{Translating into real-world applications}
The final frontier will consist in translating social contagion research to real-world social problems beyond applications to product adoption and advertisement.
Such studies could focus on inducing social contagion that intends a positive impact on society. 
Topics that citizens of a society can democratically agree are to the benefit of everyone.
For instance, researchers in Facebook have already demonstrated that it is possible to significantly increase the participation to the election by engineering the social contagion on Facebook alone~\cite{bond201261}. 
Similar campaigns may be designed and implemented for the behaviors that are relevant to public health, such as hand washing, safe sex, or vaccination. 

At the same time, it is essential to remain vigilant with respect to the other side of the coin: increasing our collective ability to detect and mitigate malicious manipulative campaigns or public shaming events. 
It has been shown that there exist ongoing efforts to manipulate public opinions through human workers, social bots, and fake news~\cite{howard2017troops, king2017chinese}. 
This type of manipulation potentially threatens the foundation of democracy in many countries across world and even the very concept of `truth'. 
Thus it will be important for researchers to ask how the study of social contagion can help us understand and improve the `post-truth' world. 
\begin{center}
	$\bullet\,\bullet\,\bullet$
\end{center}
And to the reader who has made it this far. We thank you!

\end{document}